\documentclass{appolb}
\usepackage{graphicx}

\begin{document}
\title{FISSION FRAGMENT MASS AND KINETIC ENERGY YIELDS OF FERMIUM ISOTOPES
\thanks{Presented at the Mazurian Lakes Conference on Physics, 2023, Poland.}}
\author{K.~Pomorski$^a$\footnote{Email: Krzysztof.Pomorski@umcs.pl, 
ORCID ID: 0000-0002-5557-6037}, A.~Dobrowolski$^a$, B.~Nerlo-Pomorska$^a$, M.~Warda$^a$, A.~Zdeb$^a$, J.~Bartel$^b$,
H.~Molique$^b$, C.~Schmitt$^b$, Z.~G.~Xiao$^c$, Y.~J.~Chen$^d$, L.~L.~Liu$^d$
\address{$^a$Institute of Physics, Maria Curie Sk{\l}odowska University, 
            Lublin, Poland\\
         $^b$ IPHC, University of Strasbourg, Strasbourg, France\\
         $^c$ Department of Physics, Tsinghua University, Beijing, China\\
         $^d$ China Institute of Atomic Energy, Beijing,  China}}
\maketitle
\begin{abstract}
A rapidly converging 4-dimensional Fourier shape parametrization is used to model the fission process of heavy nuclei. Potential energy landscapes are computed within the macroscopic-microscopic approach, on top of which the multi-dimensional Langevin equation is solved to describe the fission dynamics. Charge equilibration at scission and de-excitation by neutron evaporation of the primary fragments after scission is investigated. The model describes various observables, including fission-fragment mass, charge, and kinetic energy yields, as well as post-scission neutron multiplicities and, most importantly, their correlations, which are crucial to unravel the complexity of the fission process. The parameters of the dynamical model were tuned to reproduce experimental data obtained from thermal neutron-induced fission of $^{235}$U, which allows us to discuss the transition from asymmetric to symmetric fission along the Fm isotopic chain.
\end{abstract}
\vspace{1cm}

Since the discovery of the fission phenomenon in 1938, it has been commonly accepted that the most probable mass of the heaviest fragment produced in spontaneous or low-energy fission is located at $A\approx 140$. However, a series of experiments by Hulet and co-workers \cite{Hul86} have shown that this is not always the case. In the spontaneous fission of Fermium isotopes, in particular, one notices a rapid change of the fragment mass-yield systematics with growing neutron number. For light Fm isotopes, one observes a mass asymmetry of the fission fragments typical for actinides. Yet, at $A=258$, the fission fragment mass distribution changes rapidly and becomes symmetric and narrow. Also, the fragments' total kinetic energy (TKE) yield becomes significantly larger than in the lighter Fm isotopes. This new trend in the fragment distribution is also observed for No isotopes. This discovery of Hulet and co-workers constitutes a challenge for nuclear theoreticians. Several attempts have been made to explain this phenomenon, an excellent review of which is presented in the work of Albertsson et al. \cite{ACD21}. In the present paper, we will show that this rapid change of the fission yield systematics in the Fermium isotopes can be well characterized when using a new, very efficient Fourier-type parametrization \cite{SPN17} of the shapes of fissioning nuclei, combined with the WKB method to describe the penetration of the fission barrier and a Langevin type calculation \cite{PNS23} of the fragment yields.

Our ``{\it Fourier-over-Spheroid}'' (FoS) parametrization, which describes the shape of the nucleus relative to a spheroidal deformation, was recently introduced in Refs.~\cite{PNe23,PNS23} through the shape function:
\begin{equation}
  \rho_s^2(z) = \frac{R_0^2}{c} \, f\left(\frac{z-z_{\rm sh}}{z_0}\right)
\label{rhos}
\end{equation}
which describes, in cylindrical coordinates, the location of a surface point of an axially symmetric shape as function of the symmetry $z$ coordinate. In this expression, $R_0$ corresponds to the radius of the spherical nucleus having the same volume, and the parameter $c$ determines the elongation of the shape with half-axis $z_0 = c R_0$. The shift parameter $z_{\rm sh}$ guarantees that the center of mass of the shape is located at the origin of the coordinate system ($z_{\rm sh}-z_0 \leq z \leq z_{\rm sh}+z_0$). The function $f(u)$ defines a shape having half-length $c=1$:
\begin{equation}
\begin{array}{rl}
  f(u)&=1-u^2 -\left[{a_4\over 3}-{a_6\over 5}+\dots\right]\!
    \cos\left({\pi\over 2}u\right)-a_3\sin(\pi\,u) \\[0.5ex]
    &-a_4\cos\left({3\pi\over 2}u\right)
     -a_5\sin(2\pi\,u)-a_6\cos\left({5\pi\over 2}u\right)-\dots~,
\end{array}
\label{fos}
\end{equation}
where $-1\leq u \leq 1$. The first two terms in $f(u)$ describe a circle, and the third ensures volume conservation for arbitrary deformation parameters $\{a_3,\;a_4,\;\dots\}$. The parameters $a_3$ and $a_4$ allow for reflection asymmetric and necked-in shapes, with higher-order parameters $a_n, \; n \geq 5$ responsible mostly for the fission fragments deformations. Shapes breaking axial symmetry can easily be described through the shape function 
\begin{equation}
  \varrho_s^2(z,\varphi) = \rho_s^2(z) \, {1-\eta^2\over 1+\eta^2+2\eta\cos(2\varphi)} \;.
\label{rhosa}
\end{equation}
with the parameter $\eta=(b-a)/(b+a)$, where $a$ and $b$ are the half-axis of the ellipsoid obtained as the cross-section of the non-axial shape for constant $z$ (see e.g. \cite{SPN17}). The parametrization (\ref{rhosa}) is similar but more general than the $\gamma$-deformation of {\AA}ge Bohr. Equation (\ref{rhosa}) describes the same class of shapes as the original Fourier deformation parameter set of Ref.\ \cite{SPN17} but is better adapted for performing numerical calculations around the scission point of fissioning nuclei.
\begin{figure}[t!]
\includegraphics[width=\textwidth]{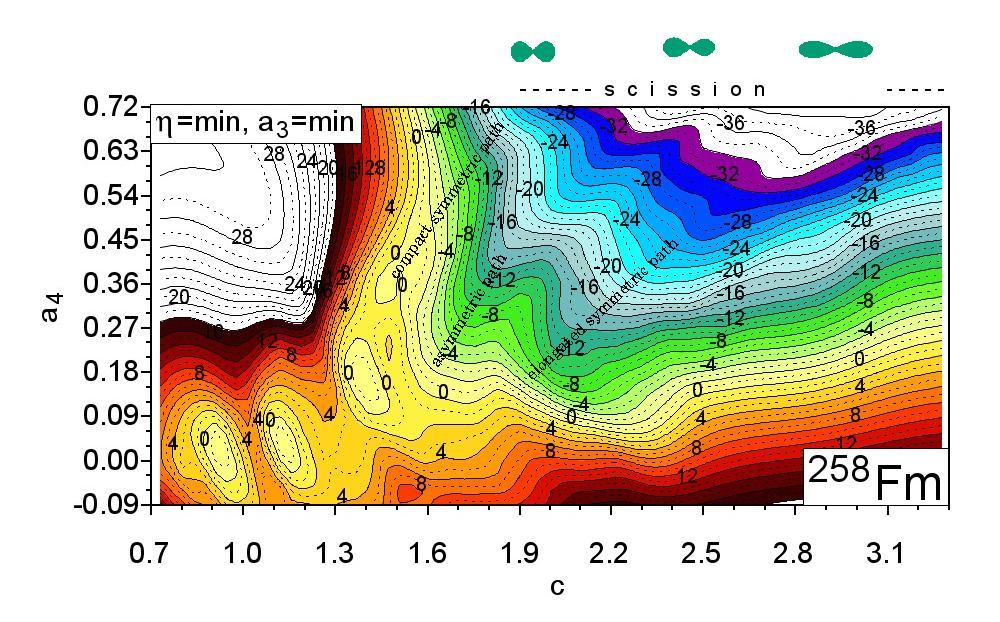}\\[-5ex]
\caption{Potential energy surface of $^{258}$Fm on the ($c,\,a_4$) plane. Each point is minimized with respect to the non-axial ($\eta$) and the reflectional ($a_3$) deformations.}
\label{PES}
\end{figure}

Using the above FoS parametrization, we have performed extensive mac\-roscopic-microscopic calculations of the PESs for even-even nuclei with $90\leq {\rm Z}\leq 122$ \cite{PNS23} in the 4D $\{\eta,c,a_3,a_4\}$ deformation space. The Lublin-Strasbourg Drop (LSD) formula \cite{PDu09} was used to evaluate the macroscopic part of the energy, while the microscopic one are obtained using the single-particle spectra of a Yukawa-folded mean-field Hamiltonian \cite{DPB16} and the Strutinsky shell and BCS pairing corrections. An example of the obtained potential energy surface (PES) is presented in Fig.~\ref{PES}, where the energy of  $^{258}$Fm (renormalized to the LSD energy of the spherical nucleus) is presented as function of the elongation $c$ and the neck-parameter $a_4$. Each point of the surface is minimized with respect to the non-axial ($\eta$) and reflectional asymmetry ($a_3$) deformations. The ground-state of $^{258}$Fm is found to be located at $c=1.13$ and $a_4=0.03$, with $a_3=\eta=0$. Interestingly, two-second saddles are found: one at $c=1.41$ and $a_4=0.27$ leading to compact symmetric fission and a second one, around 1 MeV higher, at $c=1.55$ and $a_4=0.12$, which opens a path leading both to asymmetric mass fragments, but bifurcates also into another valley characterized by a symmetric mass split with very elongated fragments. The shapes of the forming fission fragments corresponding to these three valleys are characterized on top of Fig.~\ref{PES}. The FoS parametrization scission line is found around $a_4 \approx 0.72$. 
\begin{figure}[t!]
\begin{center}
\includegraphics[width=0.5\textwidth]{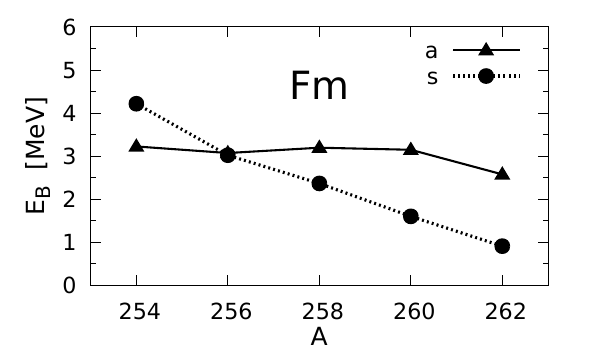}\\[-12ex]
\end{center}
\caption{Second fission barrier heights of $^{254-262}$Fm corresponding to the asymmetric (a) and symmetric (s) saddle points.}
\label{bar}
\end{figure}
\begin{figure}[t!]
\begin{center}
\includegraphics[width=0.7\textwidth]{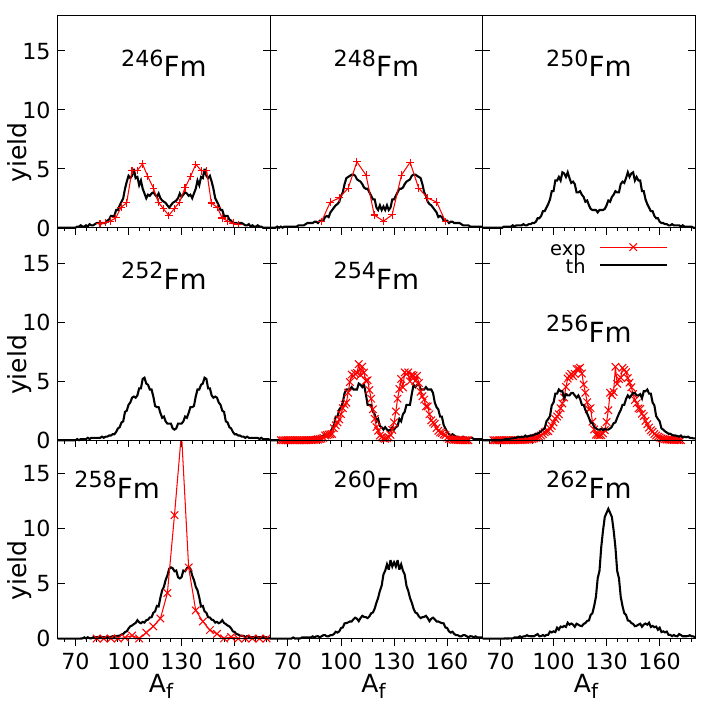}\\[-6ex]
\end{center}
\caption{Fission fragment mass yields of Fermium isotopes. The experimental data (*) are taken from Refs~\cite{Hul86,SJA16,RDB10,HWW80}.}
\label{myfm}
\end{figure}

The competition in energy between the different second saddles and subsequent fission valleys in Fm isotopes decides which fission path will be more populated. The second barrier height, defined as the energy difference between the 2$^{\rm nd}$ saddle and the ground state of the Fm isotopes, is shown in Fig.~\ref{bar}.

Using the sets of Langevin equations defined in the FoS deformation space, one obtains the mass and total kinetic energy of the fission fragments \cite{LCW21,KDN21}. We have chosen the corresponding symmetric or asymmetric exit point from the fission barrier as a starting point of the Langevin trajectories.

The final mass and TKE yields ($Y_{\rm th}$) were obtained by weighting the yields obtained for both starting points ($Y_a$ and $Y_s$) using the penetration probability ($P_a$ and $P_s$) of the corresponding fission barrier:
\begin{equation}
Y_{\rm th}(A_f)=P_a\cdot Y_a(A_f)+P_s\cdot Y_s(A_f)~.
\end{equation}
The relative probabilities $P_i$ are given through the corresponding action integral $S_i$ determined in the WKB approximation:
\begin{equation}
P_i=\exp(-S_i)/[\exp(-S_a)+\exp(-S_s)]
\end{equation} 
evaluated along the path $i$ (confer e.g.\ to Ref~\cite{PDN22}).
\begin{figure}[h!]
\begin{center}
\includegraphics[width=0.6\textwidth]{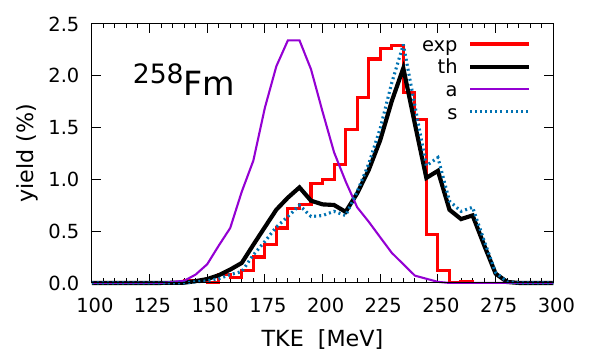}\\[-6ex]
\end{center}
\caption{Fission fragment yield (thick solid line) of $^{258}$Fm as function of the fragment TKE. The experimental data (histogram) are taken from Ref.~\cite{Hul86}. A thin solid line shows the TKE distribution corresponding to the asymmetric path, and the dotted line the one for the compact symmetric path.}
\label{Fm-tke}
\end{figure}

Our estimates of the mass yields of $^{246-262}$Fm isotopes are compared with the experimental distributions in Fig.~\ref{myfm}. 
The corresponding fission fragment TKE yield of $^{258}$Fm is shown in Fig.~\ref{Fm-tke}. The predicted yield (thick solid line) reproduces the general trend observed in the measured TKE distribution \cite{Hul86}.

\begin{figure}[t!]
\includegraphics[width=0.5\textwidth]{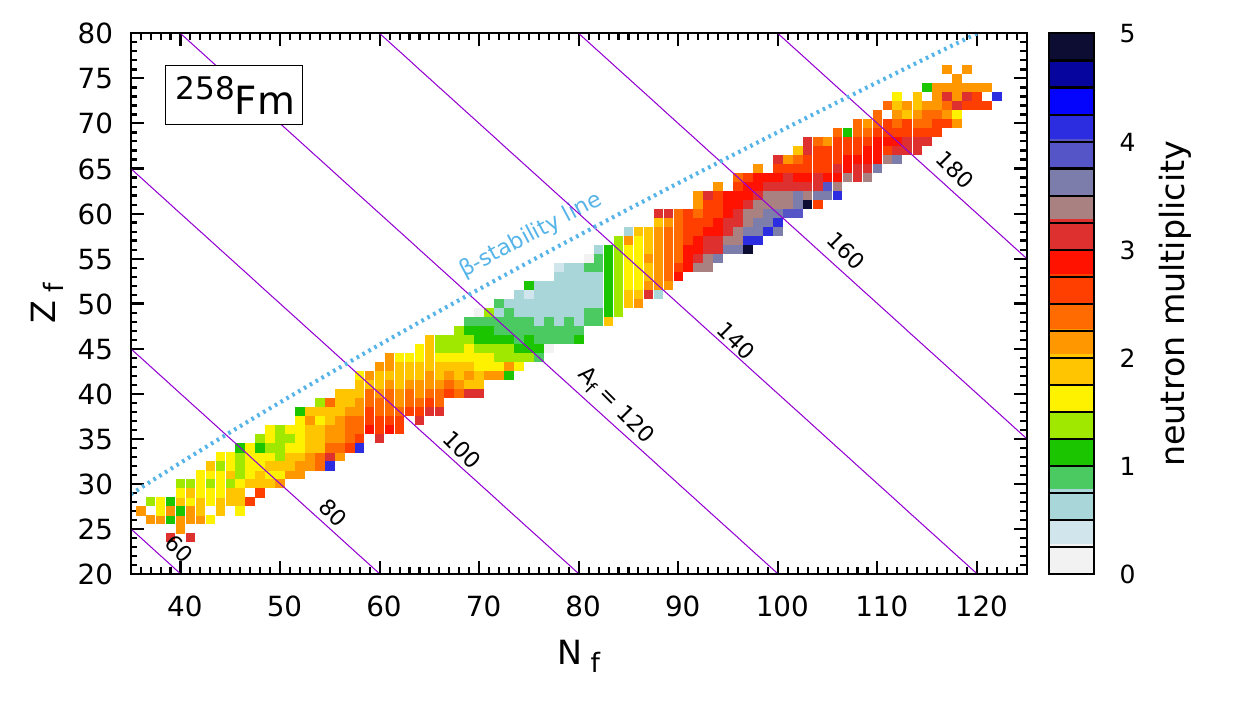}
\includegraphics[width=0.5\textwidth]{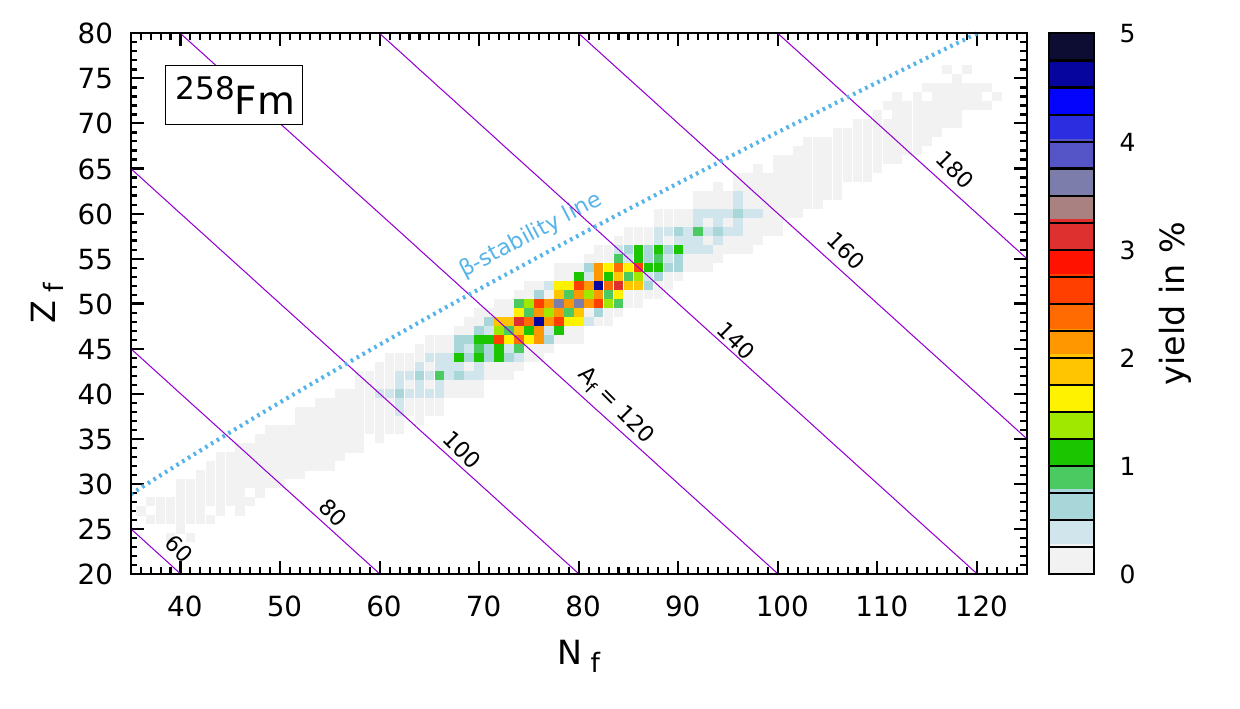}\\
\includegraphics[width=0.5\textwidth]{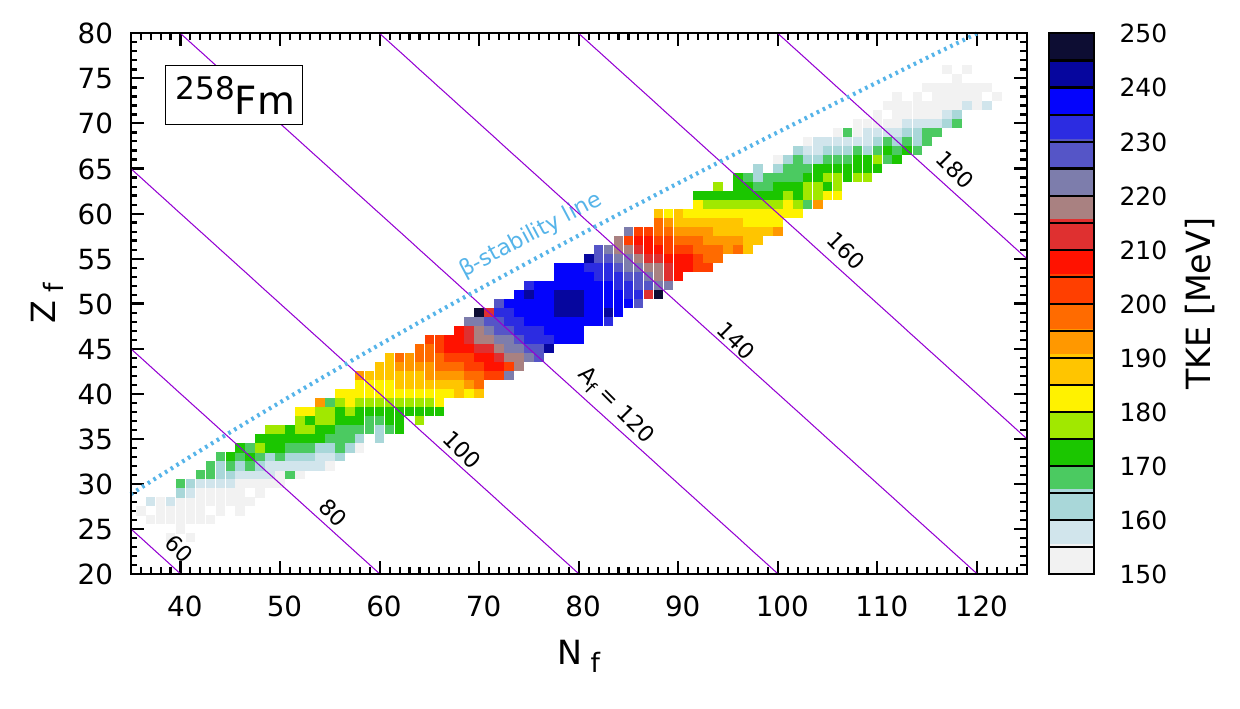}
\includegraphics[width=0.5\textwidth]{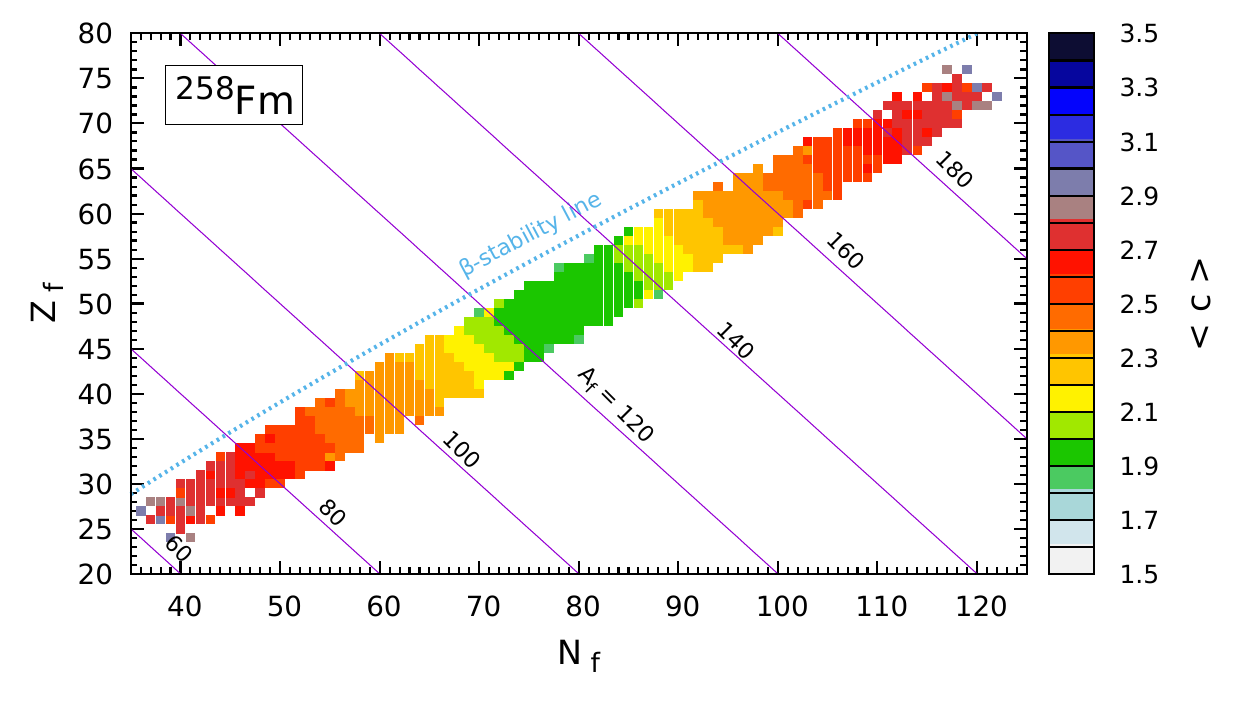}\\[-4ex]
\caption{Maps of the post-fission neutron multiplicities (top l.h.s.), fission fragment yield (top r.h.s), TKE (bottom l.h.s.), and elongation at scission (bottom r.h.s.) for $^{258}$Fm.}
\label{multi}
\end{figure}

Taking into account the charge equilibration effect, the mass and deformation, and the excitation energy of the fragments, one can estimate the post-fission neutron multiplicities as described in Ref.~\cite{PNS23}. The multiplicities of neutrons accompanying the spontaneous fission of $^{258}$Fm, the corresponding isotope and TKE yields and the elongation of nucleus at scission are shown in Fig.~\ref{multi} as function of the fragment neutron $N_f$ and proton $Z_f$ numbers, which illustrates the possibilities of our numerical code to account for the different aspects of the nuclear fission process.


\section*{Conclusions}

The following conclusions can be drawn from our investigation:\\[-4ex]
\begin{itemize}
\item The Fourier expansion of nuclear shapes offers a very effective way of describing the deformation of fissioning nuclei in the vicinity of the ground state and the scission point.
\item Potential energy surfaces are evaluated in the macro-micro model using the LSD mass formula for the liquid-drop type energy and a Yukawa-folded single-particle potential to obtain the microscopic energy correction.
\item A 3D Langevin model that couples fission, neck, and mass asymmetry modes was used to describe the fragment mass and kinetic energy yields.
\item The transition with increasing mass from asymmetric to compact symmetric fission, as observed in Fermium isotopes, is well reproduced.
\item The multiplicity of post-scission neutrons and the charge of the fission fragments are estimated within our model.
\item The influence of the inclusion of higher-multipolarity deformation parameters $a_5$ and  $a_6$ is on our agenda.
\end{itemize}
Further calculations for a wider mass and charge region are in progress.\\[4ex]

\parindent 0pt
{\bf Acknowledgements}\\[1ex]
This work was supported by the Polish National Science Centre, project No. 2018/30/Q/ST2/00185, the Natural Science Foundation of China (Grant  No.11961131010 and 11790325), and the COPIN-IN2P3 collaboration (project number 08-131) between PL-FR labs.


\end{document}